\documentstyle[epsf]{article}

\newskip\humongous \humongous=0pt plus 1000pt minus 1000pt

\newif\ifdtup

\def\oldreffmt#1{\rlap{[#1]} \hbox to 2\parindent{}}

\def\figfmt#1{\rlap{Figure {#1}} \hbox to 1in{}}

\def\beq{\begin{equation}}
\def\eeq{\end{equation}}

\relax

\font\tenbf=cmbx10
\font\tenrm=cmr10
\font\tenit=cmti10
\font\elevenbf=cmbx10 scaled\magstep 1
\font\elevenrm=cmr10 scaled\magstep 1
\font\elevenit=cmti10 scaled\magstep 1

\renewenvironment{thebibliography}[1]
 { \elevenrm
   \begin{list}{\arabic{enumi}.}
    {\usecounter{enumi} \setlength{\parsep}{0pt}
     \setlength{\itemsep}{3pt} \settowidth{\labelwidth}{#1.}
     \sloppy
    }}{\end{list}}

\def\dfone {
\begin{figure}
\vspace{4.5cm}
\vspace{-0.75cm}
\caption{\label{lf1}
Feynman diagrams for the process $e\gamma \rightarrow \nu_e q \overline{q}'$ }
\vspace{-0.5cm}
\end{figure}
 }
\def\dftwo {
\begin{figure}
\vspace{-0.4cm}
\begin{center}\mbox{
\epsfxsize=12cm
\epsffile{ocipc9210-2.pst}
 }
\end{center}
\vspace{-0.7cm}
\caption{\label{lf2}
Distributions for 500 GeV $e\gamma$\ collisions:
Solid line is S.M., dashed-dotted is $\kappa$ = 1.5,
dashed line is $\kappa$\ = 0.5,
and dotted line is $\lambda$\ = 1.0}
\vspace{-0.6cm}
\end{figure}
 }
\def\dfthree {
\begin{figure}
\vspace{-0.0cm}
\begin{center}\mbox{
\epsfxsize=12cm
\epsffile{ocipc9210-3.pst}
 }
\end{center}
\vspace{-00.7cm}
\caption{\label{lf3}
Distributions for 1 TeV $e\gamma$\ collisions: Same legend as figure 2.}
\vspace{-0.4cm}
\end{figure}
 }
\def\dffour {
\begin{figure}
\vspace{-0.5cm}
\begin{center}\mbox{
\epsfxsize=10.0cm
\epsffile{ocipc9210-4.pst}
 }
\end{center}
\vspace{-0.75cm}
\caption{\label{lf4}
a) Limits on $\kappa_\gamma$\ and $\lambda_\gamma$\
at the 95\% confidence level for a 500 GeV $e\gamma$\ collider:
Dashed line is limit based upon 4 $\theta_W$\ bins, dotted line
is based upon 4 $p_{TW}$\ bins, and solid line is the combined
measurement.
b) Same as a) except for 1 TeV $e\gamma$\ collider:
The additional vertical dash-dotted line is limit based upon cross-sectional
measurement with $M_{q\overline{q}'} >$\ 600 GeV.
 }
\vspace{-01.0cm}
\end{figure}
 }

\begin{document}
\vspace*{-1.5cm}
\begin{flushright}OCIP/C 92-10\end{flushright}
\begin{center}{{\tenbf CONSTRAINTS ON ANOMALOUS WW$\gamma$\
COUPLINGS FROM $e\gamma$\ COLLISIONS
\footnote{talk presented by K. Andrew Peterson at DPF92, Fermilab, Nov. 10-14,
1992} }
\vglue 1.0cm
{\tenrm K. ANDREW PETERSON and STEPHEN GODFREY\\}
\baselineskip=13pt
{\tenit Department of Physics, Carleton University, 1125 Colonel By Drive\\}
\baselineskip=12pt
{\tenit Ottawa, Ontario, Canada.  K1S 5B6\\}
\vglue 0.8cm
{\tenrm ABSTRACT}}
\end{center}
\vglue 0.3cm
{\rightskip=3pc
 \leftskip=3pc
 \tenrm\baselineskip=12pt
 \noindent
We study the potential for using $e\gamma$\ collisions produced by
backscattered laser photons
to investigate WW$\gamma$\ couplings.
We present results for Next Linear Collider energies
of 500 GeV and 1 TeV.
We find that where statistics allow, off W mass shell
results can be quite important, complementing on W mass
shell results from this and other studies.  It is shown
that $e\gamma$\ colliders would prove quite valuable
in the investigations of W-boson physics.
\vglue 0.6cm}
{\elevenbf\noindent 1. Introduction}
\vglue 0.2cm
\baselineskip=14pt
\elevenrm
The study of gauge boson interactions is still
unexplored territory.   To this point, our only
glimpse of these interactions is via radiative
loop corrections induced by these interactions
at low energies.  Although, high precision measurements
give us some information on the gauge boson couplings,
a direct test of them is still necessary.  The first such
direct test of triple gauge boson couplings will come at
LEP200, where sufficient $\sqrt{s}$\ energy exists for W pair production.
However, the limited available phase space means precision
tests of the three vector coupling will undoubtedly await the
arrival of a 500 GeV to 1 TeV $e^+ e^-$\ linear collider.

By limiting ourselves to  $e^+ e^-$\ collisions, we limit the information
that can be obtained.  W pair production can proceed through both
$\gamma$\ and Z exchange, and one ends up probing both the WWZ and
WW$\gamma$\ vertex simultaneously.   A way to isolate one
of these vertices is therefore needed.  An $e\gamma$\ collider provides such
a method.
There exists two possible means of creating an
$e\gamma$\ collider from a $e^+e^-$\ collider.  The first possibility
uses a combination of beamstrahlung photons and
classical bremstrahlung\cite{bla:88,hal:92}.
This possibility, however, suffers from its soft photon distribution.
Hard photon distributions may be obtained through backscattering high
intensity laser beams off the incoming electron beams \cite{gin:83}.
It is this
possibility that we will concentrate on.

To parametrize the $WW\gamma$\ vertex we use a
common parametrization of the WW$\gamma$\ vertex that
imposes C and P symmetries separately \cite{hag:87,kan:89}:
\begin{equation}
{\cal L} = - i e \left\{  ( W_{\mu\nu}^\dagger W^\mu A^\nu -
W_\mu^\dagger A_\nu W^{\mu\nu} ) + \kappa_\gamma W_\mu^\dagger W_\nu A^{\mu\nu}
+ {\frac{\lambda_\gamma}{M_W^2}} W_{\lambda\mu}^\dagger W_\nu^{\,\mu}
A^{\nu\lambda}
\right\} .
\end{equation}
In the standard model, $\kappa_\gamma = 1$\
and $\lambda_\gamma = 0$.

One may obtain model independent estimates of these anomalous
couplings in the context of a  chiral lagrangian frameword
\cite{daw:91}, where one expects $\delta\kappa_V \sim O(10^{-2})$,
and $\lambda_V \sim O(10^{-4})$.  Model specific calculations bear
these estimates out \cite{cou:87}. What this implies is
that unless there is some radical new physics, one must
be able to measure these anomalous couplings to the percent
level if one would like to see new physics.
This paper will show that this percent level
could be achieved at a
$\sqrt{s}$ = 500 GeV or 1 TeV $e^+e^-$\ collider;
the Next Linear Collider,
operating in $e\gamma$\ mode.

\vglue 0.5cm
{\elevenbf \noindent 2. Calculations and Results}
\vglue 0.4cm
In previous papers of this nature\cite{cho:91,yeh:91},
calculations were performed
using the process $e\gamma \rightarrow \nu_e W$, with the appropriate
decay widths to the observed final states.  This is a decent
approximation to the actual process, since the greatest percentage
of the cross section proceeds through a real W.  However, we feel
by making this approximation, possible valuable physics is lost.
Although the off resonance cross sections are small, the deviations
of non-standard model gauge couplings in these cross sections can
be significant.  We have therefore considered the processes
$e\gamma \rightarrow \nu_e q\overline{q}'$,
$e\gamma \rightarrow \nu_e \mu \overline{\nu}_\mu$,
and $e\gamma \rightarrow \nu_e e \overline{\nu}_e$,
which may proceed via the four diagrams of figure 1.
The process, $e\gamma \rightarrow \nu_e e \overline{\nu}_e$\ also
proceeds through a Z exchange diagram not shown here.
In what follows we will concentrate on the $q \overline{q}'$\ modes.
\dfone

Amplitudes for these processes were calculated using the CALKUL
helicity technique.  In the case of the anomalous magnetic moment,
$\kappa_{\gamma}$, plus standard model terms, these amplitudes
can be found in an earlier paper by Couture {\em et al}\cite{cou:89}.
Monte Carlo integration techniques
were then used to perform the  phase space integrations and
calculate the cross sections.  The photon distributions
are treated as structure
functions,
which are then integrated with the $e\gamma$\ cross section
to obtain our results.
The exact forms of these photon distributions, along with the required
parameters will be given in a longer paper \cite{pet:92}.

\dftwo  \dfthree
Figures 2 and 3 display some of the relevant distributions of the
$q\overline{q}'$ cross section for a $\sqrt{s}$\ of 500 GeV and 1 TeV
in the case of a backscattered photon.
Displayed are the differential cross
sections as functions of the the $q\overline{q}'$\ invariant mass,
$M_{q\overline{q}'}$, the transverse momentum of the reconstructed
W, $p_{TW}$, and the angular distribution of the reconstructed W,
$\theta_W$.  A beamline angular cut of $10^o$\ was made on the
quark directions, as well as a 5 GeV cut on $p_{TW}$.
For the $p_T$\ and angular distributions, we have
also made a cut on
$M_{q\overline{q}'}$\ of 75 $ < M_{q\overline{q}'} <$\ 85.

{}From the invariant mass
distribution we see that
off W mass shell results can be particularly useful.
Although the cross sections here are small, small deviations
in either $\lambda_\gamma$\ and $\kappa_\gamma$\ can produce
order of magnitude differences in the cross section.
The $p_T$\ distribution
also provides useful information.  Here, especially in the
high $p_T$\ regions, the non-standard model couplings deviate significantly
from the standard model values.  We will use a combination of $p_T$\ bins
and angular distribution bins to provide our best constraints on the
anomalous couplings.

To obtain constraints we consider the process
$e\gamma \rightarrow \nu_e q \overline{q}'$
and assumed a integrated luminosity of
10 $fb^{-1}$.  Our total cross sections at 500 GeV
and 1 TeV are 16.7 pb and 18.9 pb respectively, so
statistics should not be the limiting factor.
We have assumed a systematic error of 5\%
on cross sections and a systematic error of 3\% for ratios
of cross sections.  The total error is then taken to be the statistical
and systematic error combined in quadrature.
It should be emphasised that both the integrated luminosity and
the systematic errors are conservative estimates.  We expect that
more realistic estimates of these variables  will
significantly improve our constraints.

Using the error in the standard model values as the error,
we calculate the $\chi^2$\ values.
\samepage{Figure 4 a) is the resulting 95\% confidence
limits on $\kappa_\gamma$\
and $\lambda_\gamma$\ for a 500 GeV collider
based on a 4 bin $p_{TW}$ measurement and a 4 bin $\theta_W$
measurement.
One sees that both $\kappa_\gamma$\ and $\lambda_\gamma$\
are constrained to 6\%.
Figure 4 b) repeats the above analysis for a 1 TeV collider.  Figure
4 b) also includes the confidence level that is obtained through a measurement
of the cross section with invariant mass, $M_{q \overline{q}'} >$\ 600 GeV.
This is not included in the combined limits.\nopagebreak\vspace{-0.0cm}
One sees that here $\kappa_\gamma$\ is still constrained to 6\%
but $\lambda_\gamma$\ to 2\%.

}
\pagebreak
\dffour
\vglue 0.3cm
{\elevenbf \noindent 4. Conclusions \hfil}
\vglue 0.4cm

We see that an $e\gamma$\ collider operating at either
500 GeV or 1 TeV can constrain the non-standard model gauge couplings
$\kappa_\gamma$\ and $\lambda_\gamma$\ to the percent level.
These constraints represent an improvement on similiar constraints
from LEP and are comparable with those obtainable at
SSC or LHC.
We conclude that one would increase
the knowledge of triple gauge boson couplings
by running an $e^+e^-$\ accelerator in a backscattered laser $e\gamma$\ mode.

\vglue 0.3cm
{\elevenbf \noindent 5. Acknowledgements \hfil}

The authors would like to thank Dean Karlen, F. Halzen, M.C. Gonzalez-Garcia
for their helpful input.  This work was supported in part by
the Natural Sciences and Engineering Research Council of Canada.

\vglue 0.5cm
{\elevenbf\noindent 6. References \hfil}
\vglue 0.4cm

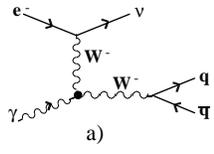
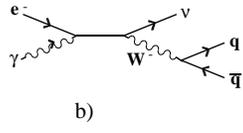
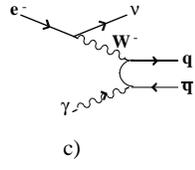
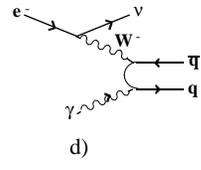
\end{document}